\documentclass[aps,amssymb,prl,preprint]{revtex4}

\begin{document}

\title{Light bending by a Coulomb field and the Aichelburg-Sexl ultraboost}

\author{ M.~V.~Kozyulin}
\affiliation{Department of physics, Novosibirsk State University, 630 090, 
Novosibirsk, Russia }

\author{ Z.~K.~Silagadze}
\affiliation{Budker Institute of Nuclear Physics and 
Novosibirsk State University, 630 090, Novosibirsk, Russia }

\begin{abstract}
We use light deflection by a Coulomb field, due to non-linear quantum
electrodynamics effects, as an opportunity for a pedagogical discussion
of the electrodynamical analog of the Aichelburg-Sexl ultraboost.
\end{abstract}

\maketitle

\section{Introduction}
Gravitational light deflection, predicted by General Relativity, is a 
fascinating phenomenon with numerous important applications in astronomy,
astrophysics and cosmology \cite{1-1,1-2,1-3}.

At first sight, there is no analogous effect in electrodynamics because
Maxwell's equations are linear and, therefore, photon does not interact with
the electromagnetic field of a alleged deflector charge. However, quantum
electrodynamical corrections bring nonlinearities in the theory \cite{1-4,
1-5}. As a result, in a external electromagnetic field, the vacuum acquires
an effective refractive index \cite{1-6,1-6P}
\begin{equation}
n\approx 1+\epsilon \vec{Q}^2,
\label{eq1-1}
\end{equation}
where 
\begin{equation}
\epsilon=\frac{a\alpha^2\hbar^3}{45m^4c^5}
\label{eq1-2}
\end{equation}
and 
\begin{equation}
\vec{Q}=\vec{\tau}\times \vec{E}+\vec{\tau}\times (\vec{\tau}\times \vec{B}).
\label{eq1-3}
\end{equation}
Here $a=14$ or $a=8$ depending on the polarization mode of the photon and
$\vec{\tau}$ is a unit vector in the direction of light propagation.

One immediate consequence of this effective refractive index is the 
deflection of light in a Coulomb field \cite{1-7,1-8}. In this paper we use 
light bending by a Coulomb field as a pedagogical tool for a discussion of 
the electrodynamical analog of the general relativistic Aichelburg-Sexl 
ultraboost \cite{1-9}.

\section{Light bending in a Coulomb field}
In geometrical optics, the light trajectory in an inhomogeneous medium is 
determined by the equation \cite{2-1,2-2}
\begin{equation}
\frac{d}{ds}\left (n\vec{\tau}\right )=\nabla n,
\label{eq2-1}
\end{equation}
where $n$ is the index of refraction and $\vec{\tau}=\frac{d\vec{r}}{ds}$ is 
the unit tangent vector to the light ray. In the Coulomb field, equations
(\ref{eq1-1}) and (\ref{eq1-3}) give
\begin{equation}
n\approx 1+\epsilon (\vec{\tau}\times \vec{E})^2,
\label{eq2-2}
\end{equation}
where 
$$\vec{E}=\frac{Ze}{4\pi r^2}\frac{\vec{r}}{r}$$
is the Coulomb field of a nucleus with electric charge $Ze$ 
(in the Heaviside-Lorentz rationalized natural unit system). For reasonable
impact parameters, the nonlinear effects are very small, $\epsilon E^2\ll 1$, 
and the index of refraction is only slightly different from unity.

It is clear from the symmetry of the problem that the light trajectory is a 
planar curve and therefore we can assume $ds=\sqrt{dx^2+dy^2}$ in Cartesian
coordinates. The tiny light deflection angle can be found as follows. From 
(\ref{eq2-1}) we get
\begin{equation}
n\frac{d\tau_y}{ds}+\tau_y\frac{dn}{ds}=(\nabla n)_y=\frac{\partial n}
{\partial y}.
\label{eq2-4}
\end{equation}
However, both $\tau_y$ and $\frac{dn}{ds}$ are small quantities (we assume
that the incident light ray was in the $x$-direction). Therefore,
in a linear approximation, (\ref{eq2-4}) can be replaced by
\begin{equation}
\frac{d\tau_y}{ds}=\frac{\partial n}{\partial y},
\label{eq2-5}
\end{equation}
and, hence,
\begin{equation}
\tau_y=\int\limits_\gamma \frac{\partial n}{\partial y}\;ds,
\label{eq2-6}
\end{equation}
where the integration contour $\gamma$ is the light trajectory. But for small
deflection angles (and note that $\frac{\partial n}{\partial y}$ in  
(\ref{eq2-6}) is very small), we can assume a rectilinear light trajectory
while calculating the integral in (\ref{eq2-6}) and, finally,
\begin{equation}
\tau_y\approx \int\limits_{-\infty}^\infty \frac{\partial n}{\partial y}\;dx.
\label{eq2-7}
\end{equation} 
In the vicinity of the rectilinear light trajectory,
$$(\vec{\tau}\times \vec{E})^2=E^2\,\frac{y^2}{r^2}=\frac{Z^2e^2}{16\pi^2}\,
\frac{y^2}{(x^2+y^2)^3}$$
and we get, on the trajectory,
\begin{equation}
\frac{\partial n}{\partial y}=
\frac{\epsilon Z^2 e^2}{8\pi^2}\left (\frac{b}{(x^2+b^2)^3}-
\frac{3b^3}{(x^2+b^2)^4}\right ),
\label{eq2-8}
\end{equation}
where $b$ is the impact parameter for the incoming light ray (that is, the 
equation of the trajectory is $y=b$). Substituting (\ref{eq2-8}) into 
(\ref{eq2-7}), we get
\begin{equation}
\tau_y=\frac{\epsilon Z^2 e^2}{8\pi^2}\int\limits_{-\infty}^\infty
\left (\frac{b}{(x^2+b^2)^3}-
\frac{3b^3}{(x^2+b^2)^4}\right )\, dx.
\label{eq2-9}
\end{equation}
The evaluation of the integral in (\ref{eq2-9}) can be facilitated by the 
following trick a la Feynman \cite{2-3} (the mathematically inclined reader 
can try to find the rigorous justification of this seemingly dubious method. 
Other interesting integration tricks can be found in \cite{2-4}). We note 
that
$$\int\limits_{-A}^A
\left (\frac{b}{(x^2+b^2)^3}-
\frac{3b^3}{(x^2+b^2)^4}\right )\, dx=\frac{1}{2}\left [ b\;\frac{\partial^2}
{\partial(b^2)^2}+b^3\;\frac{\partial^3}{\partial(b^2)^3}\right ]
\int\limits_{-A}^A \frac{dx}{x^2+b^2}.$$
Therefore,
\begin{equation}
\tau_y=\frac{\epsilon Z^2 e^2}{32\pi^2}\lim_{A\to\infty}
\left [\frac{\partial}{\partial b}
\left (\frac{1}{b}\,\frac{\partial}{\partial b}\right )+\frac{b^2}{2}
\frac{\partial}{\partial b}\left (\frac{1}{b}\,\frac{\partial}{\partial b}
\left (\frac{1}{b}\,\frac{\partial}{\partial b}\right )\right )\right ]
\frac{\arctan{\frac{A}{b}}}{b}.
\label{eq2-10}
\end{equation}
However, it is easy to see that all terms that originate from the derivatives 
of $\arctan{\frac{A}{b}}$ vanish in the limit $A\to\infty$. Therefore,
\begin{equation}
\tau_y=\frac{\epsilon Z^2 e^2}{64\pi}\left [\frac{\partial}{\partial b}
\left (\frac{1}{b}\,\frac{\partial}{\partial b}\right )+\frac{b^2}{2}
\frac{\partial}{\partial b}\left (\frac{1}{b}\,\frac{\partial}{\partial b}
\left (\frac{1}{b}\,\frac{\partial}{\partial b}\right )\right )\right ]
\frac{1}{b}=-\frac{9\epsilon Z^2 e^2}{128\pi b^4}.
\label{eq2-11}
\end{equation}
Substituting here $\epsilon$ from (\ref{eq1-2}), we get for the light 
deflection angle, in accordance with \cite{1-7},
\begin{equation}
\alpha\approx\sin{\alpha}=|\tau_y|=\frac{a\alpha^3 Z^2}{160}\,\left(
\frac{\lambda_e}{b}\right )^4,
\label{eq2-12}
\end{equation}
where we have introduced the Compton wavelength of the electron $\lambda_e=
\frac{\hbar}{mc}$. 

\section{Aichelburg-Sexl ultraboost for a Coulomb field}
What is the electromagnetic field of a massless charged particle? This is a 
classic textbook problem \cite{3-1} with elegant and interesting solution. 
Considered in a number of publications \cite{3-2,3-3,3-4,3-5,3-6,3-7,3-8,
3-9,3-10} at various levels of mathematical rigor, this problem, however, has 
been largely ignored in classical electrodynamics textbooks (the third 
edition of the Jackson's classic \cite{3-11} already has it).

It is plausible to assume that the electromagnetic field of a massless 
charged particle is a limiting case of the field of a ultrarelativistic
charged particle with finite mass. In the rest frame $S^\prime$ of a charge
$Ze$ we have the Coulomb field
\begin{equation}
\vec{E}^{\,\prime}=\frac{Ze}{4\pi r^{\prime\,2}}\,\frac{\vec{r}^{\,\prime}}
{r^\prime},\;\;\vec{B}^{\,\prime}=0,\;\; r^{\prime\,2}=x^{\prime\,2}+
y^{\prime\,2}+z^{\prime\,2}.
\label{eq3-1}
\end{equation}
In the laboratory frame $S$, where the charge moves with velocity $v$ along
the $x$-axis, the electromagnetic field is given by \cite{3-11,3-12} (we will 
assume $c=1$ for the light velocity in the rest of the paper)
\begin{eqnarray}
E_x&=& E^\prime_x, \hspace*{28mm} \; B_x= B^\prime_x, \nonumber \\
E_y&=&\gamma \left (\vec{E}^{\,\prime}-\vec{v}\times \vec{B}^{\,\prime}
\right)_y, \;\;
B_y=\gamma \left (\vec{B}^{\,\prime}+\vec{v}\times \vec{E}^{\,\prime}
\right)_y,  \nonumber \\
E_z&=&\gamma \left (\vec{E}^{\,\prime}-\vec{v}\times \vec{B}^{\,\prime}
\right)_z, \;\;
B_z=\gamma \left (\vec{B}^{\,\prime}+\vec{v}\times \vec{E}^{\,\prime}
\right)_z. 
\label{eq3-2}
\end{eqnarray}
In combination with the Lorentz transformation
\begin{equation}
x^\prime=\gamma (x-vt),\;\;\;t^\prime=\gamma (t-vx),\;\;\;y^\prime=y,\;\;\;
z^\prime=z,
\label{eq3-3}
\end{equation}
(\ref{eq3-1}) and (\ref{eq3-2}) give
\begin{equation}
E_x=\frac{Ze}{4\pi\gamma^2 R^3}\,(x-vt),\;E_y=\frac{Ze}{4\pi\gamma^2 R^3}\,y,
\;E_z=\frac{Ze}{4\pi\gamma^2 R^3}\,z,\;\vec{B}=\vec{v}\times\vec{E},
\label{eq3-4}
\end{equation}
where
\begin{equation}
R=\sqrt{(x-vt)^2+\gamma^{-2}(y^2+z^2)}.
\label{eq3-4R}
\end{equation}
We need the limiting case of (\ref{eq3-4}) when $v\to 1$. In the gravitational
case, analogous problem was considered by Aichelburg and Sexl in their seminal 
paper \cite{1-9}. Therefore, usually such a limit is called the 
Aichelburg-Sexl ultraboost.

Note that
$$\lim_{\gamma\to\infty}\frac{\gamma^{-2}}{R^3}=\left \{\begin{array}{c}
\infty,\;\;\; \mathrm{if} \; x-t=0, \\ 0,\;\;\;\; \mathrm{if} \; x-t\ne 0,
\end{array}\right .$$
but
$$\int\limits_{-\infty}^\infty \frac{\gamma}{(\gamma^2x^2+\rho^2)^{3/2}}\,dx
=-2\lim_{A\to\infty}\frac{\partial}{\partial\rho^2}\int\limits_{-A}^A
\frac{d\tau}{\sqrt{\tau^2+\rho^2}}=-4\lim_{A\to\infty}\frac{\partial}
{\partial\rho^2}\;{\rm arcsinh}\,\frac{A}{\rho}=\frac{2}{\rho^2}.$$
Therefore, we conclude that
\begin{equation}
\lim_{\gamma\to\infty}\frac{\gamma^{-2}}{R^3}=\frac{2}{y^2+z^2}\,\delta(x-t).
\label{eq3-5}
\end{equation}
Another way to obtain this result is the following one \cite{3-4}. We have
$$\frac{\gamma^{-2}}{R^3}=\frac{1}{y^2+z^2}\,\frac{\partial}{\partial x}
\left ( \frac{x-vt}{R}\right ).$$
Hence
$$\lim_{\gamma\to\infty}\frac{\gamma^{-2}}{R^3}=\frac{1}{y^2+z^2}\,
\frac{\partial}{\partial x}\lim_{v\to 1}\frac{x-vt}{R}=\frac{1}{y^2+z^2}
\,\frac{\partial}{\partial x}\left (\frac{x-t}{|x-t|}\right ).$$
But
$$\frac{x-t}{|x-t|}=2\,\theta(x-t)-1,$$
and remembering that 
$$\frac{\partial}{\partial x}\,\theta(x-t)=\delta(x-t),$$
we again get (\ref{eq3-5}). Here $\theta(x)$ is the Heaviside step function
which is unity if $x>0$ and zero if $x<0$.

In both versions of derivation of (\ref{eq3-5}), we had somewhat carelessly 
interchanged derivatives and limits. Fortunately, for generalized functions, 
and the considered limit makes sense only in the context of generalized 
functions, derivatives and limits always commute \cite{3-6}.

Using (\ref{eq3-5}) and the identity $(x-t)\,\delta(x-t)=0$, we get from
(\ref{eq3-4}) the electromagnetic field after the Aichelburg-Sexl ultraboost
\begin{equation}
E_x=0,\;\;\;E_y=\frac{Ze}{2\pi}\,\frac{y}{y^2+z^2}\,\delta(x-t),\;\;\;
E_z=\frac{Ze}{2\pi}\,\frac{z}{y^2+z^2}\,\delta(x-t),\;\;\;
\vec{B}=\vec{i}\times\vec{E},
\label{eq3-6}
\end{equation}
where $\vec{i}$ is the unit vector in the $x$-direction.

Alternatively, we can solve directly the Maxwell equations by introducing
The electromagnetic four-potential $A^\mu=(\Phi,\vec{A})$ \cite{3-5,3-7}.
In the Lorentz gauge, $\partial_\mu A^\mu=0$, the equation for the 
four-potential has the form (remember, we are using Heaviside-Lorentz units)
\begin{equation}
\Box A^\mu= J^\mu, 
\label{eq3-7}
\end{equation}
where $$\Box=\partial_\mu \partial^\mu=\frac{\partial^2}{\partial t^2}-
\nabla^2$$
is the d'Alembert operator.

For a massless charge $Ze$ which moves with the light velocity $c=1$ along 
the $x$-axis, the charge density is $\rho=Ze\,\delta(x-t)\, \delta(y)\,
\delta(z)$ which implies the following current density $J^\mu=Ze\,n^\mu
\delta(x-t)\, \delta(y)\,\delta(z)$, where $n^\mu=(1,1,0,0)$. Therefore, we
search a solution of (\ref{eq3-7}) in the form $A^\mu=A^\mu(x-t,y,z)$, so 
that $(\partial_t^2-\partial_x^2)A^\mu=0$. Substituting in (\ref{eq3-7}), we
get (here $\Delta_2=\partial_y^2+\partial_z^2$ is two-dimensional Laplacian)
\begin{equation}
\Delta_2(\Phi-A_x)=\Delta_2 A_y=\Delta_2 A_z=0,
\label{eq3-8}
\end{equation}
and
\begin{equation}
\Delta_2 \Phi=-\rho=-Ze\,\delta(x-t)\, \delta(y)\,\delta(z).
\label{eq3-9}
\end{equation}
The equations (\ref{eq3-8}) do not contain the charge $Ze$. Therefore,
we choose the solutions which are natural for a zero charge
\begin{equation}
A_y=A_z=\Phi-A_x=0.
\label{eq3-10}
\end{equation}
Note that (\ref{eq3-10}) become evident if we write the solution of 
(\ref{eq3-7}) by employing the Green's function method ($x$ and $y$ denote 
four-vectors here):
$$A^\mu(x)=\int G(x-y)J^\mu(y)\,dy.$$
As for equation (\ref{eq3-9}), it is essentially a two-dimensional 
electrostatic problem with the solution (see, for example \cite{3-13})
\begin{equation}
\Phi=-\frac{Ze}{4\pi}\,\delta(x-t)\,\ln{(y^2+z^2)}.
\label{eq3-11}
\end{equation}
It can be easily checked that, through the standard relations
$$\vec{E}=-\nabla\Phi-\partial_t \,\vec{A},\;\;\;
\vec{B}=\nabla\times\vec{A},$$
equations (\ref{eq3-10}) and (\ref{eq3-11}) reproduce the electromagnetic 
field (\ref{eq3-6}).

There is a third way to get an electromagnetic field of a massless charge, 
by considering the Aichelburg-Sexl ultraboost for the four-potential itself.
In a massive charge's rest frame $S^\prime$ we have
\begin{equation}
\Phi^\prime=\frac{Ze}{4\pi r^\prime},\;\;\;\;\vec{A}^{\,\prime}=0.
\label{eq3-12}
\end{equation}
In the laboratory frame $S$, the four-potential $A^\mu=(\Phi,\,\vec{A})$ can
be obtained by the Lorentz transformations of (\ref{eq3-12}), and we easily
find
\begin{equation}
\Phi=\frac{Ze}{4\pi R},\;\;\;A_x=\frac{Zev}{4\pi R},\;\;\;A_y=A_z=0,
\label{eq3-13}
\end{equation}
where $R$ is given by (\ref{eq3-4R}).

And here we have a problem because it is somewhat tricky to find the limit of 
(\ref{eq3-13}) when $\gamma\to\infty$ \cite{3-2,3-4}. However, the gauge 
invariance of electrodynamics comes to our rescue. Note that, if $x-t\ne 0$,
when
\begin{equation}
\lim_{v\to 1}A^\mu=\frac{Ze}{4\pi}\,\frac{n^\mu}{|x-t|},
\label{eq3-14}
\end{equation}
and this is a pure gauge giving a zero electromagnetic field in accordance
with (\ref{eq3-6}). This suggests to subtract this gauge term from 
(\ref{eq3-13}) and hope that the remaining will converge to some 
$\delta$-function when $\gamma\to\infty$. However, (\ref{eq3-14}) is singular
at $x-t=0$. Therefore, we first regularize it by considering the four-potential
\begin{equation}
A^\mu_\Lambda(x-t)=\frac{Ze}{4\pi}\,\frac{n^\mu}{R_\Lambda},
\label{eq3-15}
\end{equation}
where
\begin{equation}
R_\Lambda=\sqrt{(x-t)^2+\gamma^{-2}\Lambda^2},
\label{eq3-16}
\end{equation}
$\Lambda$ being an arbitrary parameter which just sets the regularization 
scale. Note that (\ref{eq3-15}) is also a pure gauge because
\begin{equation}
\frac{1}{R_\Lambda}=\frac{\partial}{\partial x}\ln{\left (x-t+\sqrt{(x-t)^2+
\gamma^{-2}\Lambda^2}\right)}=-\frac{\partial}{\partial t}\ln{\left (x-t+
\sqrt{(x-t)^2+\gamma^{-2}\Lambda^2}\right)}.
\label{eq3-17}
\end{equation}
 We therefore consider, instead of (\ref{eq3-13}), the following four-potential
\begin{equation}
\Phi=\frac{Ze}{4\pi}\left(\frac{1}{R}-\frac{1}{R_\Lambda}\right),\;\;\;
A_x=\frac{Ze}{4\pi}\left(\frac{1}{R}-\frac{1}{R_\Lambda}\right)-\frac{Ze}{4\pi 
R}(1-v),\;\;\;A_y=A_z=0,
\label{eq3-18}
\end{equation}
which gives the same electromagnetic field as (\ref{eq3-13}) because we have 
just subtracted a pure gauge term (\ref{eq3-15}) from (\ref{eq3-13}).

Note that
$$\lim_{v\to 1}\frac{1-v}{R}=0$$
is a well defined limit irrespective $x-t\ne 0$ or $x-t=0$ (except the 
naturally singular point $x-t=y=z=0$ where the charge resides). Therefore, 
$\lim_{v\to 1}(\Phi-A_x)=0$ and we recover (\ref{eq3-10}) if $\lim_{v\to 1}
\Phi$ has a well defined sense and we now show that this latter limit is 
indeed well defined in the sense of generalized functions. Writing
$$\frac{1}{R}=\frac{\partial}{\partial x}\ln{\left (x-vt+\sqrt{(x-vt)^2+
\gamma^{-2}(y^2+z^2)}\right)},$$
we get by combining it with (\ref{eq3-17})
\begin{equation}
\Phi=\frac{Ze}{4\pi}\,\frac{\partial}{\partial x}\ln{\frac{x-vt+
\sqrt{(x-vt)^2+\gamma^{-2}(y^2+z^2)}}{x-t+\sqrt{(x-t)^2+\gamma^{-2}
\Lambda^2}}}.
\label{eq3-19}
\end{equation}
Using that in the limit $\gamma\to\infty$ one has
$$\frac{x-t+|x-t|+\frac{\gamma^{-2}}{2}\,\frac{y^2+z^2}{|x-t|}}
{x-t+|x-t|+\frac{\gamma^{-2}}{2}\,\frac{\Lambda^2}{|x-t|}}\to\left \{
\begin{array}{c} \;\;\;\;\;1,\hspace*{5.5mm}\mathrm{if}\;\;\;x-t>0, \\ 
\frac{y^2+z^2}{\Lambda^2},\;\;\;\mathrm{if}\;\;\;x-t<0,\end{array}\right .$$
we get \cite{3-4}
\begin{equation}
\lim_{v\to 1}\Phi=\frac{Ze}{4\pi}\,\frac{\partial}{\partial x} \left [
\left (1-\theta(x-t)\right )\ln{\frac{y^2+z^2}{\Lambda^2}}\right ]=
-\frac{Ze}{4\pi}\,\delta(x-t)\,\ln{\frac{y^2+z^2}{\Lambda^2}},
\label{eq3-20}
\end{equation}
that is essentially the same result as (\ref{eq3-11}) because $\Lambda$ here
is just an irrelevant parameter setting the scale of the logarithm (we could,
of course, introduce $\Lambda$ already in  (\ref{eq3-11})).

\section{Small-angle scattering of a charged particle in a Coulomb field}
As an application of the limiting electromagnetic field of the previous 
section, let us consider the small angle scattering of a highly relativistic
charge $e^\prime$ on a heavy nucleus carrying a charge $Ze$ \cite{4-1}.

In the rest frame $S$ of the nucleus, the charge $Ze$ is at the spatial 
origin and the charge $e^\prime$ moves, before the collision, with a 
ultra-relativistic speed $v\approx 1$ in the positive x-direction, and the 
orientations of the $y$ and $z$ axes are chosen in such way that we have 
$z=b$ and $y=0$, $b$ being the impact parameter.
 
In the rest frame $S^\prime$ of the projectile charge $e^\prime$, the nucleus
appears to be traveling with the ultrarelativistic speed $v\approx 1$ in the
negative $x$-direction, while the charge $e^\prime$ is sitting at the point
$x^\prime=y^\prime=0,\,z^\prime=b$. Therefore, the electromagnetic field
of the nucleus in this frame is a plane impulsive electromagnetic wave given
by (\ref{eq3-6}) (with obvious substitution $x\to -x$ because now the wave is
traveling in the negative $x$-direction). When this impulsive wave meets the 
motionless charge $e^\prime$ at the time $t^\prime=0$, it will give the charge
$e^\prime$ a kick in the $z$-direction because the only nonzero component of 
the electric field of the electromagnetic wave, on the line $y^\prime=0,\,
z^\prime=b$, is  
$$E^\prime_z=\frac{Ze}{2\pi}\,\frac{1}{b}\,\delta(x^\prime+t^\prime).$$
Therefore, after the kick the charge $e^\prime$ acquires a small momentum
in the $z$-direction
\begin{equation}
p^\prime_x=p^\prime_y=0,\;p^\prime_z=\Delta p^\prime_z=\int\limits_{-\infty}
^\infty e^\prime E^\prime_z\,dt^\prime=\frac{Zee^\prime}{2\pi}\,\frac{1}{b}.
\label{eq4-1}
\end{equation}
Let us now return to the laboratory frame $S$ via the Lorentz transformations
\begin{equation}
p_x=\gamma(p^\prime_x+v\,E^\prime)\approx \gamma m,\;\;p_y=p^\prime_y=0,\;\;
p_z=p^\prime_z=\frac{Zee^\prime}{2\pi}\,\frac{1}{b},
\label{eq4-2}
\end{equation}
where in the first equation we have used $v\approx 1$ and $E^\prime=
\sqrt{m^2+p^{\prime\, 2}_z}\approx m$. Therefore, we get the following 
deflection angle in the laboratory frame
\begin{equation}
\alpha\approx \tan{\alpha}=\frac{p_z}{p_x}=\frac{Zee^\prime}{2\pi}\,
\frac{1}{mb\gamma}.
\label{eq4-3}
\end{equation}
To check that the result (\ref{eq4-3}) is correct, let us calculate it in the 
standard way \cite{4-2}. We have
\begin{equation}
\Delta\vec{p}=\int\limits_{-\infty}^\infty \vec{F}\,dt=\int\limits_\gamma
\vec{F}\;\frac{dl}{v},
\label{eq4-4}
\end{equation}
where the integration is along the projectile trajectory. For small angle
scattering, we can assume that the projectile moves along the straight line
while calculating the integral (\ref{eq4-4}). Therefore, for $v\approx 1$,
we get for the $z$-component of the projectile momentum after the scattering
\begin{equation}
p_z=\delta p_z=\int\limits_{-\infty}^\infty F_z\,dx=\frac{Zee^\prime}{4\pi}
\int\limits_{-\infty}^\infty \frac{b\,dx}{(x^2+b^2)^{3/2}}.
\label{eq4-5}
\end{equation}
The integral in (\ref{eq4-5}) is $2/b$ (it can be most easily calculated 
by substitution $x=b\,\sinh{\phi}$). Therefore,
$$p_z=\frac{Zee^\prime}{2\pi b}$$
and, since $p=mv\gamma\approx m\gamma$,
$$\alpha\approx\sin{\alpha}=\frac{p_z}{p}=\frac{Zee^\prime}{2\pi}\,\frac{1}
{mb\gamma},$$
which coincides with (\ref{eq4-3}).

\section{Light deflection in a Coulomb field as a Cheshire cat's smile}
If we try to describe the light bending by a Coulomb field in the manner of
previous chapter, we encounter an immediate obstacle. The refractive index
(\ref{eq2-2}) depends quadratically on the electromagnetic field strength.
However, the limiting field (\ref{eq3-6}) is proportional to the delta 
function. Therefore, while calculating the refractive index, we get the 
square of delta function which is notoriously ill-defined quantity.

On the other hand, the electromagnetic energy-momentum tensor is quadratic
in the fields too and, therefore, it also will contain squares of delta 
functions for the limiting field (\ref{eq3-6}).  This fact casts serious
doubts on the physical reality of the limit implied in (\ref{eq3-6}) 
\cite{3-6}. However, more careful analysis of physical premises of the
Aichelburg-Sexl limit indicates a way out of this dilemma.

While considering $v\to 1$ limit, if we want the nucleus energy to remain
finite, we must rescale its mass as follows \cite{1-9} $m\to \gamma^{-1}\,m$.
Would be a mass of pure electromagnetic origin, we would have $m\sim (Ze)^2$,
where $Ze$ is the nucleus charge. This suggests that the physically 
interesting Aichelburg-Sexl limit may require not only $m\to \gamma^{-1}\,m$,
but also the rescaling of the nucleus charge \cite{3-6}
\begin{equation}
(Ze)^2\to \gamma^{-1}\,(Ze)^2.
\label{eq5-1}
\end{equation}
Note that this is exactly the kind of charge rescaling used in considerations
of the Aichelburg-Sexl ultraboost for a Reissner-Nordstr\"{o}m black hole
\cite{5-1}.

Now, if we use the  rescaling (\ref{eq5-1}) in (\ref{eq3-4}), we get
\begin{equation}
E_x^2=\left (\frac{Ze}{4\pi} \right )^2\,\frac{\gamma^{-5}}{R^6}\,(x-vt)^2,
\;\;\; E_y^2=\left (\frac{Ze}{4\pi} \right )^2\,\frac{\gamma^{-5}}{R^6}\;y^2,
\;\;\; E_z^2=\left (\frac{Ze}{4\pi} \right )^2\,\frac{\gamma^{-5}}{R^6}\;z^2,
\label{eq5-2}
\end{equation}
and we should consider the limit of (\ref{eq5-2}) when $v\to 1$. Note that
$$(x-vt)^2\,\frac{\gamma^{-5}}{R^6}=\frac{\gamma^{-5}}{R^6}\,\left [(x-t)^2+
(1-v)\,t\,[2x-t(1+v)]\right ],$$
and the second term leads to the well defined limit (except the singular 
point $x-t=y=z=0$)
 
$$\lim_{v\to 1}\,(1-v)\,\frac{\gamma^{-5}}{R^6}=0.$$
Indeed, this is evident, if $x-t\ne 0$. But if $x-t=0$, we have
$$ \lim_{v\to 1}\,(1-v)\,\frac{\gamma^{-5}}{R^6}= \lim_{v\to 1}\,
(1-v)\,\frac{\gamma^{-5}}{\gamma^{-6}\,(y^2+z^2)^3}= \lim_{v\to 1}\,
\frac{1}{(y^2+z^2)^3}\,\sqrt{\frac{1-v}{1+v}}=0.$$
Next, we have to consider the limit
$$\lim_{\gamma\to\infty}\frac{\gamma^{-5}}{R^6}=\left \{\begin{array}{c}
\infty,\;\;\; \mathrm{if} \; x-t=0, \\ 0,\;\;\;\; \mathrm{if} \; x-t\ne 0.
\end{array}\right .$$
But 
$$\int\limits_{-\infty}^{\infty}\frac{\gamma^{-5}\, dx}{[x^2+\gamma^{-2}\,
\rho^2]^3}=\int\limits_{-\infty}^{\infty}\frac{dx}{(x^2+\rho^2)^3}=
\frac{3\pi}{8\rho^5}.$$
Most easily, this integral can be calculated as follows
$$\int\limits_{-\infty}^{\infty}\frac{dx}{(x^2+\rho^2)^3}=\lim_{A\to\infty}
\,\frac{1}{2}\,\frac{\partial^2}{\partial (\rho^2)^2}\int\limits_{-A}^A
\frac{dx}{x^2+\rho^2}=\lim_{A\to\infty}\frac{\partial^2}{\partial (\rho^2)^2}
\left(\frac{1}{\rho}\,\arctan{\frac{A}{\rho}}\right )=\frac{\pi}{2}\,
\frac{\partial^2}{\partial (\rho^2)^2}\,\frac{1}{\rho}.$$
Therefore, we see that \cite{3-6}
\begin{equation}
\lim_{\gamma\to\infty}\frac{\gamma^{-5}}{R^6}=\frac{3\pi}{8(y^2+z^2)^{5/2}}
\,\delta(x-t),
\label{eq5-3}
\end{equation}
and since $(x-t)^2\,\delta(x-t)=0$, we get the following limiting field 
squares
\begin{equation}
E_x^2=0,\;\;E_y^2=\left (\frac{Ze}{4\pi} \right )^2\,\frac{3\pi}{8}\,
\frac{y^2}{(y^2+z^2)^{5/2}}\,\delta(x-t),\;\; E_z^2=\left (\frac{Ze}{4\pi} 
\right )^2\,\frac{3\pi}{8}\,\frac{z^2}{(y^2+z^2)^{5/2}}\,\delta(x-t).
\label{eq5-4}
\end{equation}
The fields themselves, however, tend to zero under such  Aichelburg-Sexl
limit and we are left with the strange situation which is as bizarre as 
the Cheshire cat's smile \cite{5-2}. The cat (electromagnetic field) 
disappears but its grin (electromagnetic effects quadratic in the fields) 
remains.

Nevertheless, this Cheshire cat's smile, the limiting field squares 
(\ref{eq5-4}) can be used to re-derive the light deflection formula 
(\ref{eq2-12}) in the Coulomb field.

Suppose in the laboratory frame $S$ a light ray propagates in the direction 
$y=b,\,z=0$ and encounters a motionless nucleus of charge $Ze$ situated at 
the spatial origin.

In the ultrarelativistic frame $S^\prime$, which moves with the velocity
$v\approx 1$ in the same direction as the incident light ray, the contracted
electromagnetic field of the nucleus induces an effective index of 
refraction only in a thin layer moving with the speed $-v$. According the 
relations (\ref{eq1-1}) and (\ref{eq1-3}), when this layer meets the incident 
light ray, the index of refraction at the encounter point equals to (for some 
time, we will not use the primed notations, although we assume that we are in 
the frame $S^\prime$)
\begin{equation}
n=1+4\epsilon\,E^2,
\label{eq5-5}
\end{equation}
Because for large $\gamma$ the contracted electromagnetic field of the 
nucleus looks like a plane impulsive electromagnetic wave in which $E^2=B^2$, 
$\vec{E}\perp\vec{B}\perp \vec{\tau}$ and, since the wave and the photon are 
in a head-on collision, $\vec{E}\times\vec{B}\parallel (-\vec{\tau})$.   

We cannot replace the square of the nucleus electric field in (\ref{eq5-5})
by the limiting square field (\ref{eq5-4}) because the limit implies the
rescaling of the charge while in our case the charge is not actually rescaled. 
However, if $\gamma$ is large, the field square of the rescaled charge 
$(Ze)^2\gamma^{-1}$ will be close to the limiting square field. Therefore, we 
can replace $E^2$ in (\ref{eq5-5}) by $\gamma$ times the limiting square field 
and the refractive index on the photon's trajectory will take the form
\begin{equation}
n=1+4\epsilon\gamma\left(\frac{Ze}{4\pi}\right )^2\,\frac{3\pi}{8}\,
\frac{\delta(x+t)}{y^3}.
\label{eq5-6}
\end{equation}
As before, we get 
$$\tau_y\approx \int\limits_{-\infty}^\infty \frac{\partial n}{\partial y}\;
dx$$
and, while calculating the integral, we can assume the unperturbed photon
trajectory $x=t,\,y=b,\,z=0$, so that $\delta(x+t)=\delta(2x)=\frac{1}{2}
\delta(x)$. Thanks to the delta function, the integration is elementary and
we get
$$\tau_y\approx -\frac{9\pi}{4}\,\epsilon\,\gamma\left(\frac{Ze}{4\pi}
\right )^2\frac{1}{b^4}.$$
Therefore, in the $S^\prime$ system, the deflection angle of the photon is
\begin{equation}
\alpha^\prime\approx \sin{\alpha^\prime}=|\tau_y| \approx \frac{9\pi}{4}\,
\epsilon\,\gamma\left(\frac{Ze}{4\pi}\right )^2\frac{1}{b^4}.
\label{eq5-7}
\end{equation}
In the laboratory frame $S$, the deflection angle can be obtained via 
the aberration formula (see, for example, \cite{5-3}) which follows from the 
relativistic velocity addition law. Namely, when $v\approx 1$, we get
\begin{equation}
\sin{\alpha}=\lim_{v\to 1}\,\frac{\sin{\alpha^\prime}}{\gamma(1+
v\cos{\alpha^\prime})}=\frac{1}{\gamma}\,\frac{\sin{\alpha^\prime}}{1+
\cos{\alpha^\prime}}=\frac{1}{\gamma}\,\tan{\frac{\alpha^\prime}{2}}.
\label{eq5-8}
\end{equation}
Therefore,
\begin{equation}
\alpha\approx \frac{\alpha^\prime}{2\gamma}=\frac{9\epsilon Z^2e^2}{128\pi 
b^4},
\label{eq5-9}
\end{equation}
which is exactly the result implied by (\ref{eq2-11}).

\section{Concluding remarks}
We believe the material presented above will be useful for students just 
beginning their study of classical electrodynamics. While, on the one hand,
it is simple enough to follow with limited mathematical background, on the 
other hand, it illustrates some difficulties of using generalized functions 
in nonlinear physical theories like general relativity \cite{6-1}.

Aichelburg-Sexl ultrarelativistic limit of the Coulomb field is subtle.
We have a surprising result that the electrodynamics allows as a limit a
massless uncharged particle which creates no electromagnetic field, but
has a nonzero electromagnetic energy-momentum tensor \cite{3-6}, and thus 
induces electromagnetic light deflection. Physically this situation may seem
unsatisfactory, but mathematically the Aichelburg-Sexl limit is perfectly 
well defined \cite{3-9,6-1}. We have seen in the previous section that,
when appropriately used, this limit can produce physically reasonable 
results.
 
An interesting question remains whether a massless charge can really exist in 
nature. Up to now, no massless elementary particle with nonzero electric 
charge was ever found experimentally. It was argued that massless electric 
charges cannot exist in nature as they are completely locally screened in 
the process of formation \cite{6-2}. However, that screening occurs only 
at very large distances and meanwhile the massless charge, born in the hard
collisional process, may interact with electromagnetic field \cite{6-3}. 
We feel that, although  massless charged particles are undoubtedly peculiar 
objects \cite{6-4}, the final word has not been yet said on the delicate 
issue whether they really exist in nature.

\end{document}